\documentclass[12pt,preprint]{aastex}
\shorttitle{The Young Galaxy Holmberg~{\sc ix}}
\shortauthors{Sabbi et al.}
\begin{document}
\title{HOLMBERG~{\small IX}: THE NEAREST YOUNG GALAXY\altaffilmark{1}} 
\author{E.~Sabbi,\altaffilmark{2,3,5} J.S.~Gallagher,\altaffilmark{4} L.J.\ 
Smith,\altaffilmark{2,5,6} D.F.\ de Mello,\altaffilmark{7,8,9} \& M.~Mountain\altaffilmark{2}}
\email{sabbi@stsci.edu}
\altaffiltext{1}{Based on observations with the NASA/ESA \textit{Hubble
    Space Telescope}, obtained at the Space Telescope Science
    Institute, which is operated by AURA, Inc., under NASA contract
    NAS5--26555.} 
    \altaffiltext{2}{Space Telescope Science Institute, 3700 San Martin Drive,
Baltimore, MD USA }
\altaffiltext{3}{Astronomisches Rechen-Institut, Zentrum f\"ur Astronomie der Universit\"{a}t Heidelberg,
Heidelberg, Germany}
\altaffiltext{4}{University of Wisconsin, Madison, WI, USA }
\altaffiltext{5}{European Space Agency, Research and Scientific Support Department, Baltimore, MD 21218, USA }
\altaffiltext{6}{University College London, London, UK }
\altaffiltext{7}{Observational Cosmology Laboratory, Code 665, Goddard Space Flight Center, Greenbelt, MD 20771, USA }
\altaffiltext{8}{Catholic University of America, Washington, DC 20064, USA }
\altaffiltext{9}{Johns Hopkins University, Baltimore, MD 21218, USA}
\begin{abstract}
Deep images taken with the Wide Field Channel of the Advanced Camera for Surveys on board the \textit{Hubble Space Telescope} provide the basis for study the resolved stellar population of the M81 companion dwarf irregular galaxy Holmberg~{\sc ix}. Based on color-magnitude diagrams the stellar population toward Holmberg~{\sc ix} contains numerous stars with ages of $\la\,$200~Myr as well as older red giant stars. By charting the spatial distribution of the red giant stars and considering their inferred metallicities, we concluded that most of these older stars are associated with M81 or its tidal debris. At least 20\% of the stellar mass in Holmberg~{\sc ix}was produced in the last $\sim\,$200~Myr, giving it the youngest stellar populations if any nearby galaxy. The location of Holmberg~{\sc ix}, its high gas content, and its youthful stellar population suggests that it is a tidal dwarf galaxy, perhaps formed during the last close passage of M82 around M81.
\end{abstract}

%*
\section{Introduction} \label{intro}

\citet{zwicky56} presented the possibility that new stellar systems might form in the tidal debris around interacting galaxies. This concept has since received observational support from the detection of condensations of gas and newly formed stars in the tidal tails of interacting galaxies \citep[e.g.,][]{schwiezer78}. In some cases these structures have internal kinematics that are indicative of gravitational binding and they thus can be considered as dwarf galaxies in formation \citep[e.g.,][]{hibbard94,duc00,weilbacher00,mundell04}. The long term fate of these tidal dwarf galaxies (TDGs), however, remains unclear, although theory \citep[e.g.,][]{bournaud06} supports the possibility that TDGs separate from other tidal debris and become relatively long-lived dwarfs.

TDGs form in dynamically cool tidal tails and may favor gas rich regions \citep[e.g.,][]{wetzstein07}. In these cases the TDG stellar populations would consist of a mix of pre-existing stars from tidally disrupted materials combined with a prominent ``new generation'' of young stars produced as gas collects in the TDG. Young TDGs may stand out due to their large young stellar content; however, no TDG candidates have been found in the Local Group (LG). On the other hand the major post-formation interactions in the LG appear to have been associated with M31 and are themselves quite old, so any TDG could be difficult to distinguish and may not have survived. However if the conventional wisdom concerning a lack of dark matter in galaxy disks holds, then TDGs should contain little dark matter. Few if any nearby dwarf galaxies are candidates for satisfying this condition \citep{hunter00}.

In contrast to the Local Group, the nearby \citep[3.6~Mpc,][]{freedman94} M81 galaxy group hosts a major ongoing interaction involving M81, M82, and NGC~3077. This interaction produced extensive H{\sc i} arms that link the three galaxies \citep[e.g.,][]{gottesman75,hulst79,appleton81,yun94} and which are potential sites for TDG formation \citep[e.g.,][]{boyce01}. In particular H{\sc i} knots with optical counterparts have been identified within the tidal bridges, the most prominent to the east/north-east of M81 being Holmberg~{\sc ix}, a dwarf galaxy close to M81 catalogued by \citet{Holmberg1969}, and Arp's Loop, a region located along the bridge between M81 and M82 discovered by \citet{Arp1965}. 

We have begun a study of young stellar populations in the M81 galaxy group with the overall aim of elucidating star formation modes in tidal debris. Combining multi-wavelength observations of the stellar features in the H{\sc i} tidal bridge connecting M81 and M82, we identified eight young star-forming regions in Arp's Loop. The FUV luminosities of these objects are modest and typical of small clusters and of associations of O and B~stars. We suggested that these young stars recently formed in gas that was tidally stripped from M81 and M82 but is not part of a gravitationally bound galaxy \citep{demello08}. 

The stellar content of Holmberg~{\sc ix} was recently investigated by \citet{makarova02}. While their \textit{HST} Wide Field Planetary Camera~2 (WFPC2) observations of Holmberg~{\sc ix} were not deep enough to unambiguously rule out the presence of old stars, they noted ``...no clear signs of an RGB [red giant branch] in the color magnitude diagram of this object.''  This conclusion was reached despite overall optical properties that have led to Holmberg~{\sc ix} being considered a dwarf irregular galaxy by most investigators on the basis of its structure and kinematics.

In this {\it Letter} we analyze deep archival \textit{Hubble Space Telescope} (\textit{HST}) images of Holmberg~{\sc ix} obtained with the Wide Field Channel (WFC) of the Advanced Camera for Surveys (ACS). The data are $\sim\,$2.5~mag deeper than those obtained by \citet{makarova02}, and readily detect stars along the upper RGB, allowing us to set strong limits on the presence of old stars in Holmberg~{\sc ix}. We show that this galaxy contains a predominantly young stellar population, consistent with it being a recently formed TDG, as suggested by \citet{makarova02}. 

\section{Observations} \label{obs}
We retrieved deep broadband F555W and F814W ACS/WFC images from the Multimission Archive at STScI (MAST) of the dwarf galaxy Holmberg~{\sc ix} (P.I.\ E.D.\ Skillman, GO-10605). The dataset consists of eight 1192~s dithered exposures in both the F555W and F814W filters. The data were processed through the standard Space Telescope Science Institute ACS calibration pipeline CALACS and, for each filter, all the exposures were co-added using the MULTIDRIZZLE package \citep{koekemoer02}. For each filter the total exposure time is 4768~s and the images cover an area of $200\arcsec \times 200\arcsec$ corresponding, at a distance for Holmberg~{\sc ix} of 3.6~Mpc, to $3.5\times 3.5$~kpc. The color-combined image of the data is presented in Plate~1.

The photometric reduction has been performed with the DAOPHOT package within the IRAF\footnote{IRAF is distributed by the National Optical Astronomy Observatory, which is operated by AURA, Inc., under cooperative agreement with the National Science Foundation.} environment. Stars were independently detected in each filter using the DAOFIND routine, with a detection threshold set at $4\sigma$ above the local background level. Their fluxes were measured by aperture photometry using an aperture size of $0.15\arcsec$. We then performed PSF-fitting photometry to refine the photometric measurements of the individual sources. To take into account the spatial variations in the core width and shape of the PSF \citep{krist03,sirianni05}, we computed a spatially-variable PSF using 
$\sim\,$180 isolated and moderately bright stars, uniformly distributed over the entire region. We transformed the instrumental magnitudes to the \textit{HST} VEGAMAG system by converting the individual stellar magnitudes to an aperture radius of $0.5\arcsec$ and applying the zero points listed in \citet{sirianni05}. 

We applied selection criteria to our catalogs based on the shape of the objects, with the aim of distinguishing bona-fide single stars from extended, blended or spurious objects. To accomplish this, we considered the DAOPHOT $\chi^2$ and sharpness parameters: $\chi^2$ gives the ratio of the observed pixel-to-pixel scatter in the fit residuals to the expected scatter calculated from a predictive model based on the measured detector features, while sharpness sets the intrinsic angular size of the objects. Only objects with $\chi^2<3$ and $-0.6<\mathrm{sharpness}< 0.6$ in the F555W filter and $\chi^2<4$ and $-0.5<\mathrm{sharpness}< 0.5$ in the F814W filter have been retained. We found these values to be the best for rejecting spurious and extended objects, without also eliminating the bright stars. By inspecting the rejected objects, we recognized several candidate star clusters (i.e., fairly round but extended objects) and background galaxies. The final catalog contains 23,182 stars.
%*
\section{RESULTS}

\begin{figure}
\epsscale{.9}
\plotone{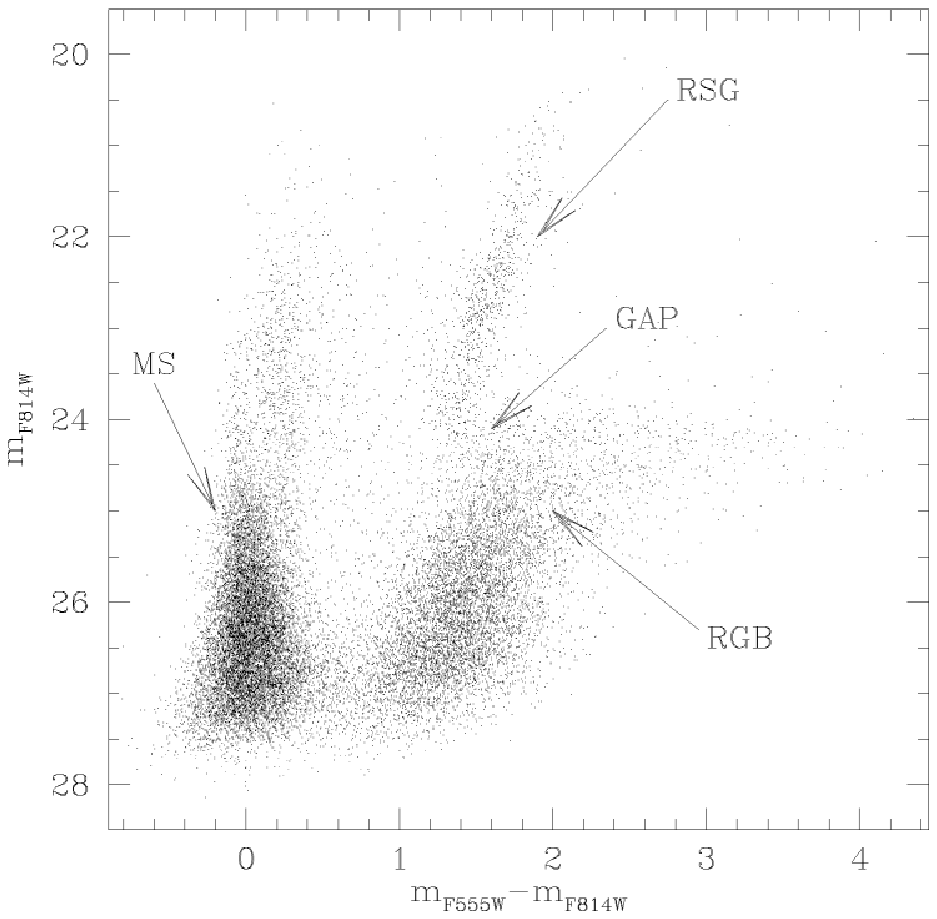}
\caption{\label{f:cmdA} $m_\mathrm{F814W}$ vs.\ $m_\mathrm{F555W}-m_\mathrm{F814W}$ CMD of the stars measured in the Holmberg~{\sc ix} region and selected with the criteria described in the text. The various stellar evolutionary phases are indicated.}
\end{figure}

Figure~\ref{f:cmdA} shows the $m_{\rm F814W}$ versus $m_{\rm F555W}-m_{\rm F814W}$ color-magnitude diagram (CMD) of Holmberg~{\sc ix}. The young stellar population is represented by well-defined blue and red plumes. The blue plume is located at $m_{\rm F555W}-m_{\rm F814W}\simeq 0.0$ with the brightest stars at $m_{\rm F814W}\simeq 20.5$ and is composed of upper main sequence (MS) stars as well as stars along the hot edge of the core helium burning blue loop phase. The quality of the photometry is such that the gap between the MS and the blue edge of the blue loop phase is clearly visible. The red plume is at $1.2\le m_{\rm F555W}-m_{\rm F814W}\le 2.0$ and extends between $20.0\le m_{\rm F814W}<24.0$. It is populated by red supergiants (RSGs) at the brighter magnitudes that define the red side of the blue loop, and by asymptotic giant branch (AGB) stars at fainter luminosities.

The rich concentration of stars fainter than $m_{\rm F814W}>24.0$, with colors redder than $m_{\rm F555W}-m_{\rm F814W}\ge1$ corresponds to low-mass, old stars (age $\ge 1$~Gyr) in the RGB evolutionary phase.  The region of 
Holmberg~{\sc ix} then contains stars covering a considerable range in age.

An interesting feature in the CMD is the lack of stars with $m_\mathrm{F814W}\le 24.0$ and colors in the range of $1\le m_\mathrm{F555W}-m_\mathrm{F814W}\le 1.5$ (marked as the ``Gap'' in Fig.~1). This shortage in stellar density at the base of the red plume indicates a prolonged quiescent star formation occurred about $\le\,$1~Gyr ago. Evidently vigorous star formation started again recently to produce the large young population seen along the blue MS and both sides of the blue loop. 

Spectroscopic observations of H{\sc ii} regions in Holmberg~{\sc ix} indicate a metallicity for the ionized gas of $Z\simeq 0.0076$ \citep{makarova02}. Therefore we used $Z=0.008$ Padova isochrones \citep{salasnich00} to derive the ages of the two stellar components identified in the Holmberg~{\sc ix} region (Fig.~2). From the structure of the RGB, the older stellar population covers a likely range in age of $\sim\,$1 to 12~Gyr, as well as having a span in metallicity that reaches to relatively high values. 

A second major episode of star formation evidently started $\simeq\!200$~Myr in the past, when M81 and M82 experienced the their nearest approach \citep[as derived from dynamical simulations by][]{yu99}, suggesting that this event could be the origin of Holmberg~{\sc ix} as a gravitationally bound entity.  As can be seen from the presence of H{\sc ii} regions, active star formation continues into the present epoch.

We next consider whether the RGB and young stars are both part of Holmberg~{\sc ix}. This is accomplished by comparing spatial distributions of the older and younger stellar populations visible in the ACS image. RGB stars with $26.0<m_\mathrm{F814W}<24.0$ and $1.0<m_\mathrm{F555W}-m_\mathrm{F814W}<3.0$ trace the old stellar population. MS and blue loop stars, the blue plume, with $26.0<m_\mathrm{F814W}<24.0$ and $-0.3<m_\mathrm{F555W}-m_\mathrm{F814W}<0.4$ chart the spatial distribution of the younger stellar populations. 

Inspection of Figure~\ref{mappe} (\textit{top panels}) reveals that the blue plume and RGB stars have very different spatial distributions. Blue plume stars (\textit{top left panel}) are strongly concentrated in the center of Holmberg~{\sc ix}, as normally observed in a star forming dwarf galaxy. The density of RGB stars (\textit{top right panel}) peaks at the left bottom corner of the image (the position closest to the M81 galaxy), and rapidly decreases toward the image center. This is not the typical spatial distribution for the old stellar population in a dwarf galaxy, which is usually symmetrically distributed around the center of the dwarf. This effect is highlighted in Figure~\ref{mappe} (\textit{lower panels}), which presents variations in stellar counts in a grid along the diagonal across the image extending from pixel (0;0) to pixel (4500;4500). Each grid element has dimensions $\simeq 150 \times150$ pixels$^2$, which corresponds to $\sim\!130\times130$~pc$^2$, at the distance of Holmberg~{\sc ix}. We performed a similar analysis on Arp's Loop \citep[see Fig.~8 in][]{demello08}, but found that the spatial distributions of the young and old stars are closely correlated. 

Experiments with more than 10$^6$ artificial stars following the procedure described in \citet{sabbi07} tested the quality of our photometry and the completeness of our data. These experiments demonstrate that the differences in the spatial distributions of blue plume and RGB stars is due neither to crowding nor incompleteness in the central region of Holmberg~{\sc ix}.  At $m_\mathrm{F814W}\le 25.7$ and $m_\mathrm{F555W} - m_\mathrm{F814W}= 1.5$, the photometry remains complete at the 90\% level (Fig.~\ref{f:cmdA}). This conclusion can be visually confirmed from the inspection of Plate~1, where a multitude of background galaxies is easily distinguishable even through the center of Holmberg~{\sc ix} where we have the highest stellar density. 

The majority of RGB stars have spatial distributions consistent with them being an extended component of M81. This also would explain the presence of a relatively metal-rich RGB stellar population in a dwarf irregular galaxy, which normally would have a more vertical RGB structure with blue colors associated with low metallicity older stars. These results reinforce the findings of \citet{makarova02}.  Holmberg~{\sc ix} is impressively dominated by stars with ages of $\la\,$200~Myr, but is it free of any traces of an older stellar populations? If Holmberg~{\sc ix} is a TDG formed in the recent M81--M82--NGC~3077 interaction, then its older stars should have come from one of the interacting systems while its young stars formed on site.

\section{DISCUSSION}

The lack of an obvious concentration of old stars associated with Holmberg~{\sc ix} is consistent with its being a TDG that formed out of a mixture of gas and stars from the disks of interacting systems in the M81 group. Since this material comes from the outer parts of galaxies, we expect it to initially have had a high gas-to-star mass ratio and low star formation rate. Alternatively Holmberg~{\sc ix} could be an old dwarf galaxy whose evolutionary path was modified through interactions. 
In particular the flattening in RGB star counts (Fig.~\ref{mappe}---right panel) may indicate that while the M81 halo dominates the lower left corner of the observed region, the RGB stars detected in the right upper could be associated with Holmberg~{\sc ix}.

To test this hypothesis we assumed that {\em all} the RGB stars detected in the right upper corner (e.g., between $\sim\,$2,000 and $\sim\,$4,000 pixels) are part of the Holmberg~{\sc ix} old stellar population. In this region we identified 447 RGB stars between $24.3<m_\mathrm{F814W}<25.7$ and $1.0<m_\mathrm{ F555W}-m_\mathrm{F814W}<2.6$. If, as usually observed in dwarf galaxies, the old stellar population is uniformly distributed, then up to $\sim\!2000$ RGB stars in the CMD could belong to Holmberg~{\sc ix}.  This is a conservative upper limit since we assumed no contamination from M81 in selecting our Holmberg~{\sc ix} RGB sample. 

We applied the synthetic CMD method of \citet{tosi91} to make a first assessment of the total numbers of young and old stars formed in Holmberg~{\sc ix}.  Under the assumption of a continuous SF between 1 and 12~Gyr ago, with a Salpeter mass function, the presence of 2000~RGB stars in the observed magnitude range requires a star formation rate of $\leq 5.5\times 10^{-4}\, M_\odot$\, yr$^{-1}$. The total mass in older stars for this model is $M_\star\leq 6\times 10^6\, M_\odot$.

Similarly we counted 1237 blue loop stars between $19.8<m_\mathrm{F814W}<24.5$ and $-0.3<m_\mathrm{F555W}-m_\mathrm{F814W}<2.7$. Assuming that the galaxy has constantly formed stars in the last 200 Myr, we derive a recent $\mathrm{SFR}\approx 8.1\times10^{-3}\, M_\odot$\, yr$^{-1}$, a factor 15 times higher than the lifetime average assuming that all of the observed younger stars formed in Holmberg~{\sc ix}. The corresponding young stellar mass is $M_\star\simeq 1.6\times 10^6\,M_\odot$. 

While we cannot exclude that Holmberg~{\sc ix} is an old dwarf galaxy, it then would have an extreme ratio of young-to-old stars. \emph{No matter what its origin, Holmberg~{\sc ix} has the youngest mean stellar population age of any nearby galaxy}. Similarly it also has an unusually high ratio of gas-to-stellar mass. We find M(H{\sc i})~$=3.3\times 10^8\, M_\odot$ from \citet[][on-line Table A.1]{swaters02}, which implies a remarkable M(H{\sc i})/M$_\star > 40$. Although dwarf galaxies with this degree of gas richness are known, it is extraordinary to find such an object apparently in close proximity to a giant spiral.

Given the current data the most viable possibility is that Holmberg~{\sc ix} is a young TDG assembled from gravitational collapse of gas and stars stripped off during the interaction of M81 with M82. It is likely that the majority of the old stars in the Holmberg~{\sc ix} region belong to the extended halo of M81 or is tidal debris not associated with Holmberg~{\sc ix}. This model receives further support from the location of Holmberg~{\sc ix} along one of the main H{\sc i} tidal arms, close to M81. Further tests of this model can be made by checking for significant dark matter in Holmberg~{\sc ix} and through more detailed studies of the age distribution of its stellar populations.

\section{CONCLUSIONS}
 
We analyzed archival \textit{HST}/WFC/ACS images of the region around the dwarf galaxy Holmberg~{\sc ix}. The resulting CMD clearly shows that this galaxy experienced an intense episode of star formation in the last $\sim\,$200~Myr. Although a prominent RGB is present in this field, the spatial distribution of the stars, and in most cases their relatively high metallicities as judged from colors, are consistent with a projected large old stellar population contribution from the halo or disk of M81. However, a slight excess of RGB stars in the region of Holmberg~{\sc ix} opens the possibility that a low mass, $M_\star\leq 6\times 10^6\, M_\odot$, old stellar component could be present in this dwarf galaxy. The best proof that Holmberg~{\sc ix} not a TDG would come from evidence for a dark matter halo.

Whatever its origin, Holmberg~{\sc ix} is a low-density stellar system, with most of its baryonic mass in the form of gas. Whether it remains gravitationally bound will depend on a variety of factors, including the fate of the gas and presence or absence of dark matter.  If Holmberg~{\sc ix} dissolves near its present location, then in a few hundred million years its stars should begin forming tidal streams similar to those observed in M31 \citep{ibata05}.

While we cannot strictly exclude that Holmberg~{\sc ix} is an old dwarf galaxy, the high ratio of gas to stellar mass would then be highly peculiar for a system located near a giant spiral. On the other hand all of the observed properties of Holmberg~{\sc ix} can be understood in the context of a TDG that formed in tidal debris $\sim\,$200~Myr in the past. In particular the gap in star formation activity at $\approx\,$1~Gyr is consistent with this galaxy having become an active star forming and likely gravitationally bound entity about 200~Myr ago, near the time of the closest M81--M82 approach, as is its location along a tidal arm.

\acknowledgments

We thank Evan Skillman for having proposed to obtain these ACS WFC observations ands the STScI Hubble Heritage team for
making Figure 4. We thank Monica Tosi and Francesca Annibali for usefull discussions and suggestions. E.S. was founded by STScI GO grant GO-1208. JSG appreciates research support for this project from the University of Wisconsin Graduate School. DFdM was founded by STScI grant-44185.

\pagebreak

\begin{figure}
\epsscale{.9}
\plotone{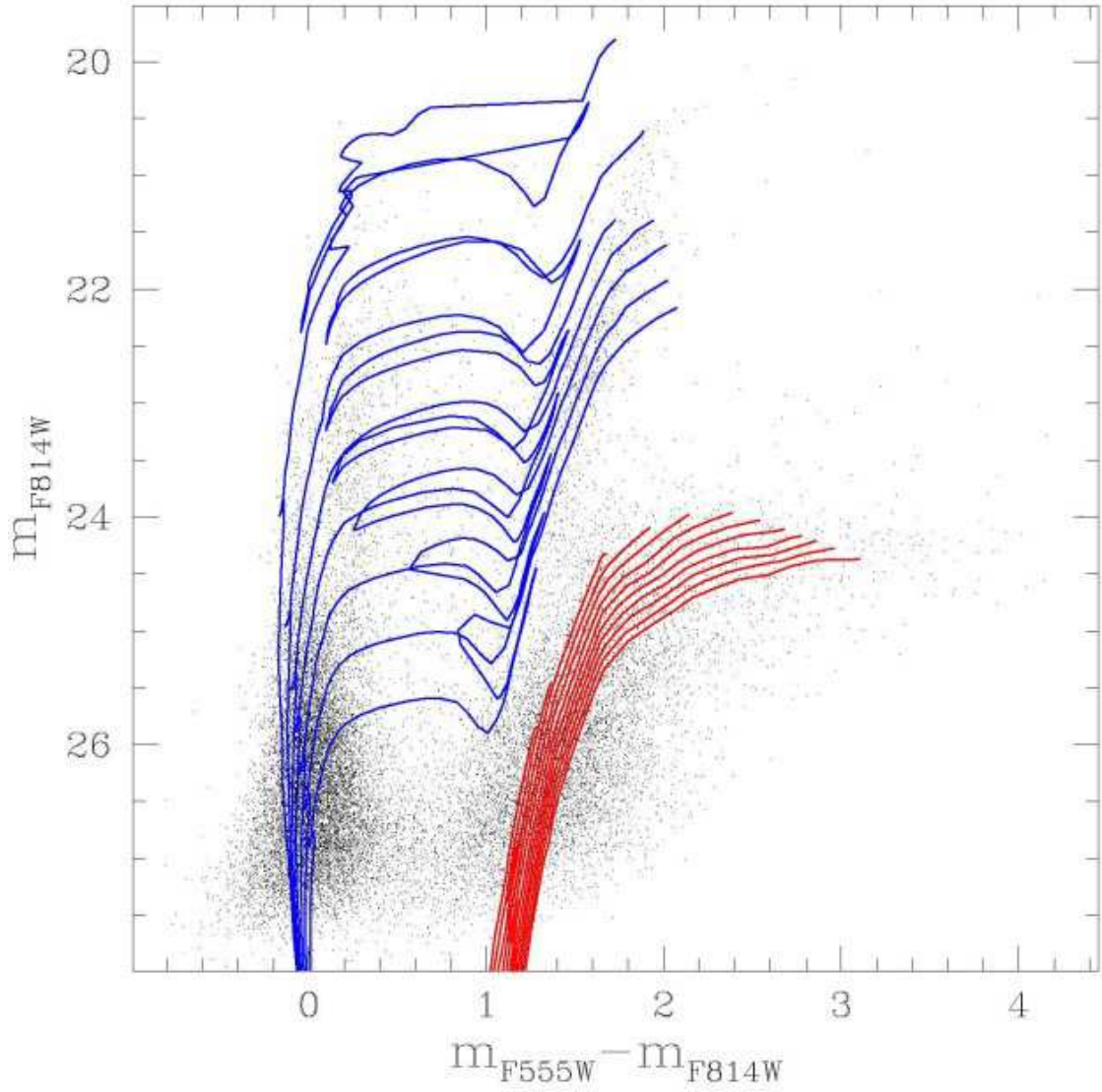}
\caption{\label{f:tracksa} Same as Figure~\ref{f:cmdA} but with superimposed $Z=0.008$ Padua evolutionary isochrones \citep{salasnich00}. Blue isochrones cover the age range between 200 and 10~Myr, while isochrones in the age range between 1 and 12~Gyr are shown in red.}
\end{figure}

\begin{figure}
\epsscale{.9}
\plotone{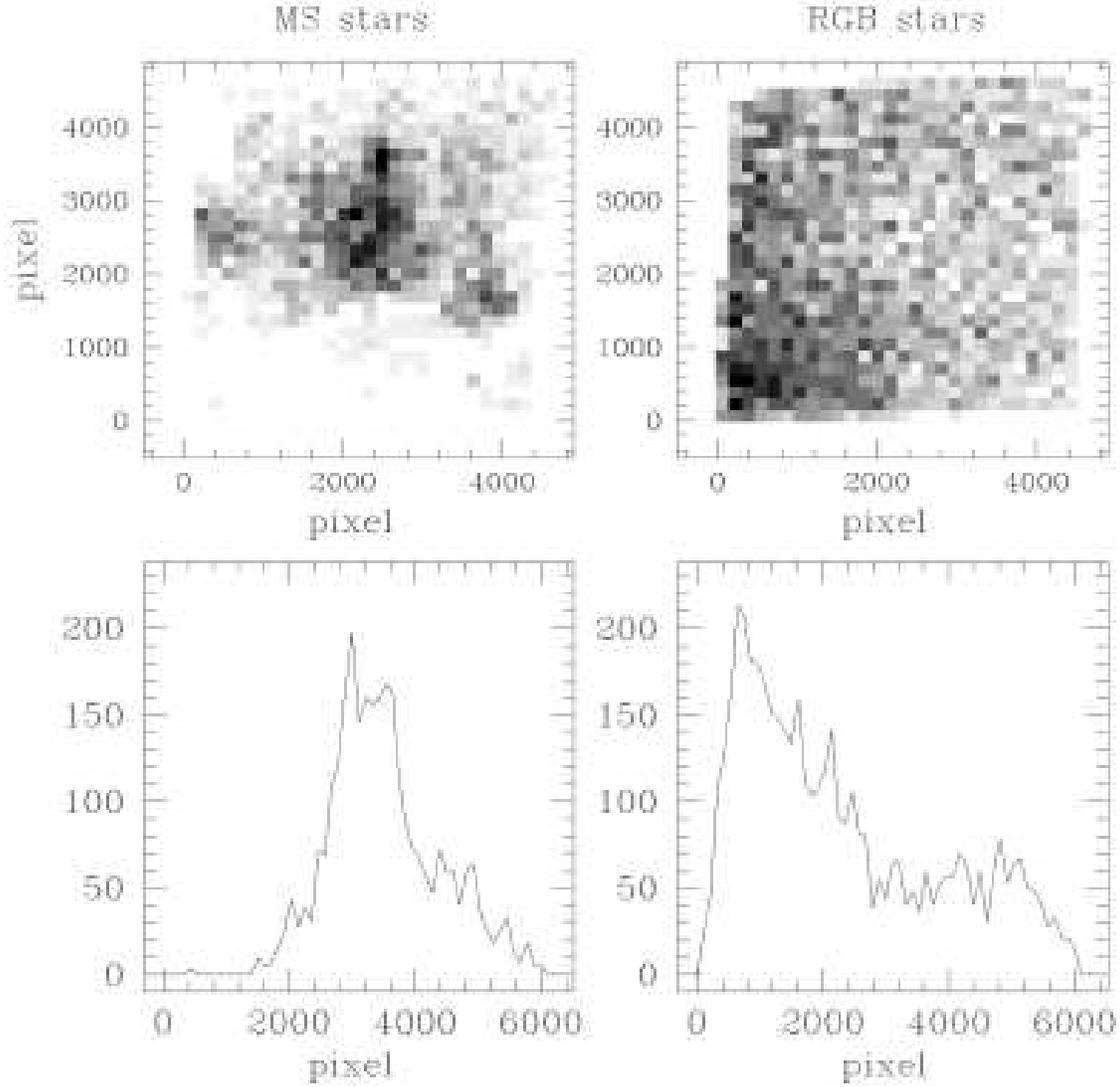}
\caption{\label{mappe} \textit{Upper panels}: Spatial stellar density distributions for MS (\textit{to the left}) and RGB (\textit{to the right}) stars. North is down-right, East is up-right. \textit{Lower panels}: Variations in stellar counts along the image diagonal from pixel 0;0 to pixel 4500;4500 for MS (\textit{to the left}) and RGB (\textit{to the right}) stars.}
\end{figure}

\begin{figure}
%\figurenum{4}
%\centering
\epsscale{1.0}
\plotone{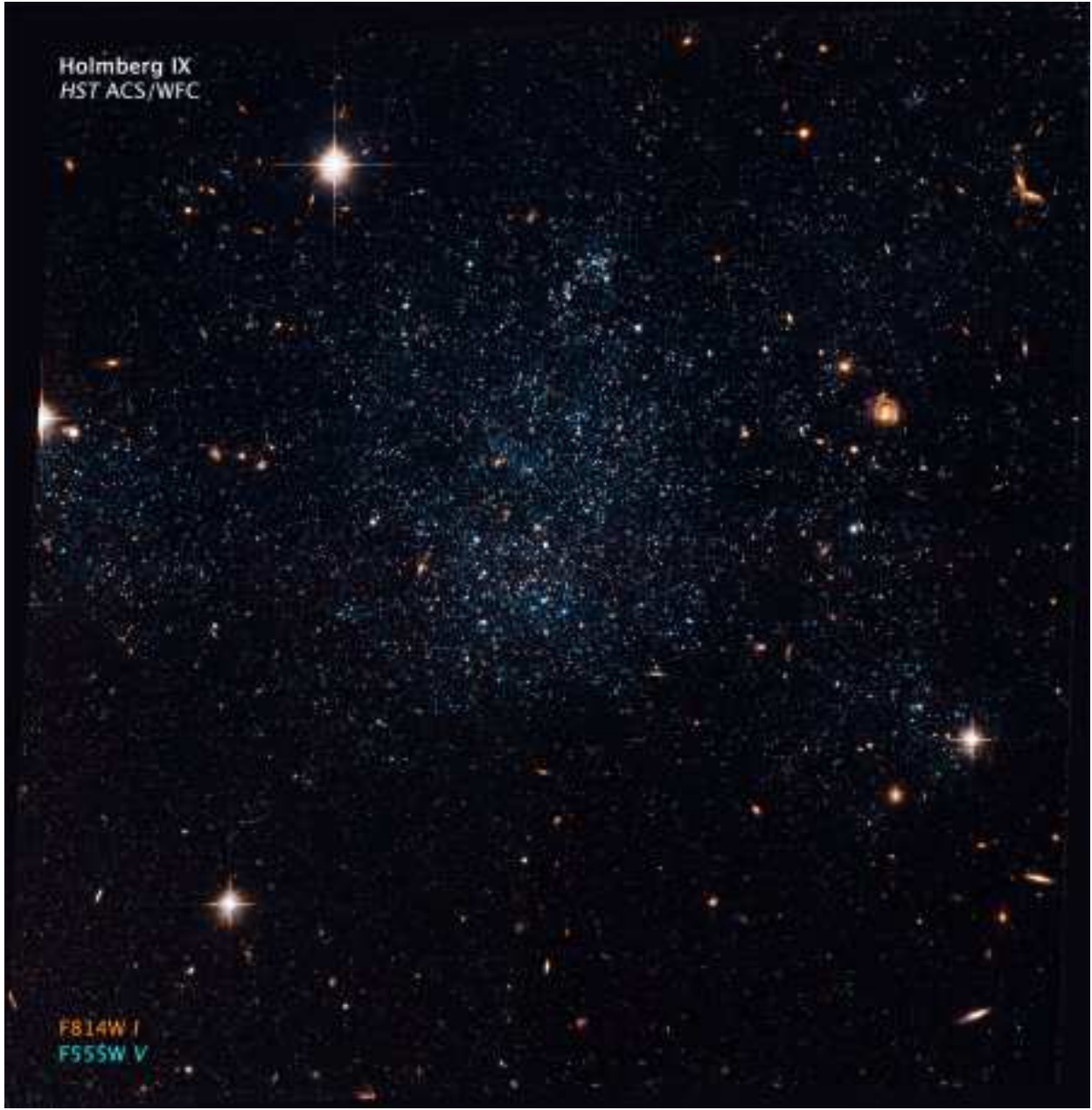}
%\resizebox{16.7cm}{!}{\includegraphics*[1,1][1020,1020]{f4.eps}}
\caption{ACS/WFC color composite image of Holmberg~{\sc ix} created using the F555W and F814W filters. The field of view is $200\arcsec \times 200\arcsec$, corresponding to $\sim\!3.5\times3.5\,\mathrm{kpc}^2$ at the distance of Holmberg~{\sc ix}. North is down-right, East is up-right.}
\label{fig:b1}
\end{figure}

\end{document}